\documentclass[%
 reprint,
 amsmath,amssymb,
 pra,
]{revtex4-1}

\pdfoutput=1

\usepackage{graphicx}% Include figure files
\usepackage{dcolumn}% Align table columns on decimal point
\usepackage{bm}% bold math
\usepackage{amsmath}
\usepackage{esvect}
\usepackage{esdiff}
\usepackage{todonotes}
\usepackage{placeins}
\usepackage{afterpage}
\DeclareMathOperator{\dd}{\mathrm{d}}
\DeclareMathOperator{\eps}{\varepsilon}

\newcommand{\eq}[1]{
\begin{equation}\begin{aligned}
#1
\end{aligned}\end{equation}}

\begin{document}

\preprint{APS/123-QED}

\title{Temporal quadratic solitons and their interaction with dispersive waves in Lithium Niobate nano-waveguides}% Force line breaks with \\

\author{William R. Rowe}
 \email{W.R.Rowe@bath.ac.uk}
\author{Dmitry V. Skryabin}
\author{Andrey V. Gorbach}%
 \email{A.Gorbach@bath.ac.uk}
\affiliation{ Centre for Photonics and Photonic Materials, Department of Physics, University of Bath, Bath BA2 7AY, United Kingdom
}%

\date{\today}% It is always \today, today,
             %  but any date may be explicitly specified

\begin{abstract}

We present a model of soliton propagation in waveguides with quadratic nonlinearity. Criteria for solitons to exist in such waveguides are developed and two example nano-waveguide structures are simulated as proof of concept. Interactions between quadratic solitons and dispersive waves are analysed giving predictions closely matching soliton propagation simulations. The example structures are found to support five different regimes of soliton and quasi-soliton existence. Pulse propagation in these example waveguides has been simulated confirming the possibility of soliton generation at experimentally accessible powers. Simulations of multi-soliton generation, Cherenkov radiation and quasi-solitons with opposite signs of dispersion in the fundamental and second harmonic are also presented here.

%\begin{description}
%\item[Usage]
%Secondary publications and information retrieval purposes.
%\item[PACS numbers]
%May be entered using the \verb+\pacs{#1}+ command.
%\item[Structure]
%You may use the \texttt{description} environment to structure your abstract;
%use the optional argument of the \verb+\item+ command to give the category of each %item. 
%\end{description}
\end{abstract}

\pacs{Valid PACS appear here}% PACS, the Physics and Astronomy
                             % Classification Scheme.
%\keywords{Suggested keywords}%Use showkeys class option if keyword
                              %display desired
\maketitle

%\tableofcontents

\section{Introduction}
\label{s:intro}

Temporal solitons are an important class of solitary waves well known in the context of nonlinear optics \cite{Kivshar2003OpticalCrystals}. Perhaps the most studied type of temporal solitons are cubic solitons, whereby self-action due to cubic ($\chi^{(3)}$ or Kerr) nonlinearity (typically self-focusing) balances dispersion, hence leading to formation of stable self-localized pulses of light propagating in optical waveguides or fibres \cite{Agrawal2013}. Effective self-action can also occur in quadratic ($\chi^{(2)}$) nonlinear media due to cascaded interactions between the fundamental frequency (FF) and second harmonic (SH) \cite{DeSalvo1992Self-focusingKTP}. In the limit of a large phase mismatch between FF and SH, the so-called cascading limit \cite{DeSalvo1992Self-focusingKTP, Buryak2002}, the system of coupled equations for FF and SH components can be reduced to an effective Kerr-type system. Notably, the sign of such effective Kerr interaction is controlled by the sign of phase mismatch. This makes the parameter space of existence of such cascaded $\chi^{(2)}$ temporal solitons to be considerably wider than in native $\chi^{(3)}$ systems \cite{Buryak2002}. However existence of quadratic temporal solitons away from the cascading limit requires a more complicated balance between nonlinear interaction, dispersion, and walk-off due to the mismatch of group velocities of the two co-propagating FF and SH pulses.

Interactions between solitons and small amplitude dispersive waves is a well-known generic mechanism of frequency conversion in Kerr media, and plays a crucial role in supercontinuum generation in optical fibres \cite{Skryabin2010ColloquiumSupercontinuum}. The corresponding theory is developed for Kerr solitons \cite{Skryabin2005TheoryFibers}, and recently was extended to cases including $\chi^{(2)}$ as a perturbation to Kerr solitons \cite{Zhou2017ParametricallyMixing} and $\chi^{(2)}$ solitons in the cascaded limit \cite{Bache2010OpticalGeneration, Zhou2015DispersiveCrystal}. Importantly this previous work does not include any predictions for dispersive waves emitted from the SH component in $\chi^{(2)}$ solitons.

Recent advancements in the fabrication of Lithium Niobate nano-waveguides \cite{Poberaj2012LithiumDevices} has reignited interest in this well characterised $\chi^{(2)}$ material \cite{Toney2015LithiumPhotonics}. The small mode size in these nano-waveguides enhances nonlinearity \cite{Poberaj2012LithiumDevices}, therefore reducing required peak powers to achieve efficient nonlinear interactions. Their strong guidance also provides geometrically tuneable dispersion allowing direct phase matching between modes \cite{Chen2018ModalPhotonics, Gorbach2015Microfiber-LithiumChip}, as well as considerable reduction of group velocity mismatch between FF and SH modes within wide frequency ranges \cite{Cai2018HighlyNano-waveguides}. Continued research has seen the loss in these structures fall as low as 0.027dB/cm \cite{Wu2018LongRoughness}, further improving the prospects of these structures for practical application. These LN nano-waveguides therefore provide novel opportunity for $\chi^{(2)}$ soliton research.

In this work a model of temporal quadratic solitons in $\chi^{(2)}$ waveguides is developed. Criteria for the existence of localised soliton solutions based on waveguide parameters are derived. The model is then extended to include the interaction of the soliton with dispersive waves. Two examples of nano-waveguide structures are simulated and found to support soliton and quasi-soliton existence in the normal dispersion regime, at experimentally attainable powers. Simulations of pulse propagation in the example nano-waveguide structures are presented and validate predictions of soliton existence and the frequency of resonant radiation.  This work develops theory of important frequency conversion phenomena with potential to enhance understanding of broadband supercontinuum generation in $\chi^{(2)}$ waveguides.

\section{Model} 
\label{s:model}
We consider a generic $\chi^{(2)}$ waveguide with a fixed cross section in the Cartesian $x$-$y$ plane and invariant along $z$, the propagation direction. We assume two pulsed light fields are excited in different modes of the waveguide at a frequency $\omega_f$ (fundamental field, FF) and its second harmonic (SH) $\omega_s=2\omega_f$ with envelope functions $U_f(z,t)$ and $U_s(z,t)$ respectively.  
The propagation constant, $\beta_m(\omega)$, for waveguide mode $m$, is related to the effective refractive index of that mode, $n_{e\!f\!f,m}$, by 
\eq{\label{e:beta_neff}
\beta_m(\omega) = \omega n_{e\!f\!f,m}(\omega)/ c 
}
where $m=f$ and $m=s$ label the waveguide mode chosen for the FF and SH respectively. The $j^\mathrm{th}$ derivative of $\beta_m$ with respect to frequency is therefore 
\eq{\label{e:beta_m_neff}
    \beta_{mj} =  \frac{\dd^j \! \beta_{m}}{\dd \omega^j}  = \frac{1}{c}\left[  j\frac{\dd^{j-1} n_{e\!f\!f,m} }{\dd \omega^{j-1} } + \omega \frac{\dd^j n_{e\!f\!f,m} }{ \dd \omega^j} \right]
}

The evolution of the field envelopes is described by:
\eq{ \label{e:model}
    i \partial_\xi U_f &=  -D_f(i\partial_\tau) U_f - U_s U_f^* e^{i\kappa \xi} , \\
i \partial_\xi U_s &= -D_s(i\partial_\tau) U_s - \frac{U_f^2}{2} 
e^{-i\kappa \xi},
}
where $t$ has been shifted to move with the FF pulse group velocity (GV), $v_f = \beta^{-1}_{f1}|_{\omega_f}$ and normalised by the approximate pulse duration, $t_0$, giving $\tau = (t-z/v_f)/t_0$. $z$ has been normalised by the dispersion length $z_d = 2t_0^2/|\beta_{f2}|$ to give $\xi=z/z_d$. 
The phase mismatch, $\kappa = \Delta \beta z_d$, where  $\Delta \beta = \beta_s(2\omega_f) - 2\beta_f(\omega_f)$ (Note that in literature $\kappa$ is often defined with the opposite sign). From the definition of $\beta_m$ in Eq. \eqref{e:beta_neff} it follows that
\eq{\label{e:phase-matching}
    \kappa  = 2  z_d \omega_f [n_{e\!f\!f,s}(2 \omega_f) - n_{e\!f\!f, f}(\omega_f) ].
}
%$\partial_{a}$ denotes the partial derivative with respect to any variable $a$. 

The dispersion operators are found by taking a Taylor series, 
\eq{
D_f(i\partial_\tau) &= -\sum_{j=2}^{+\infty} r_j[i \partial_\tau]^j,\\
D_s(i\partial_\tau) &= +s_1 [i\partial_\tau] - \sum_{j=2}^{+\infty} s_j [i \partial_\tau]^j ,
}
The walk-off parameter $s_1=z_d/z_w$ where $z_w = t_0/(v_s^{-1} - v_f^{-1})$ is the walk-off length where the SH GV, $v_s = \beta_{s1}^{-1}|_{2 \omega_f}$. The remaining dispersion coefficients are,
 \eq{\label{e:r_and_s}
 r_j &= - \frac{ z_d }{ t_0^{j} \, j! } \beta_{fj} ,\\
 s_j &= - \frac{ z_d }{ t_0^{j} \, j! } \beta_{sj},
 }
 for integers $j \ge 2$. It should be noted that $r_2=\pm1$.
The dispersion of each mode at frequency detuning $\delta = (\omega - \omega_f) t_0 = (\omega - \omega_s) t_0$ is therefore given by 
\eq{
D_f(\delta) &= -\sum_{j=2}^{+\infty} r_j \delta^j,\\
D_s(\delta) &= s_1 \delta - \sum_{j=2}^{+\infty} s_j  \delta^j ,
}

Fields are scaled such that $U_f = \sqrt{2}\rho_2 z_d A_f$ and $U_s = \rho_2 z_d A_s$ where $|A_{f,s}|^2$ are intensities in Watts. The effective non-linear coefficient \cite{Gorbach2015Microfiber-LithiumChip}, 
\eq{
\rho_2 = \frac{ \eps_0 \omega_f}{4 \sqrt{N_{s}}N_f}\iint_{n} \vv{\boldsymbol{e}}_{s} \left[  \hat{\chi}_2 \vdots \vv{\boldsymbol{e}}_f^2 \right] \mathrm{d}A_{n},
}
where ${\vv{\boldsymbol{e}}}_{f,s}$ are the electric field profiles of the chosen FF and SH waveguide modes respectively. $\hat{\chi}_2$ is the second-order nonlinear tensor of the $\chi^{(2)}$ material in the waveguide and $A_n$ is the cross section of the $\chi^{(2)}$ material, over which the integral is performed.  The normalisation factors, $N_m = [1/4]\iint_{w}[\vv{\boldsymbol{e}}_m \times \vv{\boldsymbol{h}}_m^*] + [\vv{\boldsymbol{e}}_m^* \times \vv{\boldsymbol{h}}_m] \mathrm{d}A_{w}$, where $\vv{\boldsymbol{h}}_m$ is the magnetic field profile for the mode $m$ and $A_w$ is the cross section of the whole waveguide,  over which the integral is performed (not only the $\chi^{(2)}$ material).

\section{ Soliton theory}
\label{s:sol_analysis}

\subsection{Tail analysis}

We first consider soliton solutions of the system in Eq. \eqref{e:model} with second-order dispersion only such that:

\eq{\label{e:dispersion}
    D_f(\delta) =&\, - r_2 \delta^2, \\
    D_s(\delta) =&\, s_1 \delta - s_2 \delta^2.}

Solitons are sought in the form of localized pulses co-propagating with a common group velocity $\nu$:
\eq{\label{e:sol_sub}
    U_f &= W_f(\eta) e^{i\mu\xi},\\
    U_s &= W_s(\eta) e^{i[2\mu - \kappa] \xi},
}
 where $W_f$ and $W_s$ are the soliton field profiles. The transverse coordinate, $\eta = \tau - \nu \xi$, moves with the soliton velocity $\nu$, and $\mu$ is the nonlinear correction to propagation constant. $\mu$ and $\nu$ are the soliton family parameters. Substituting into Eq. \ref{e:model} and requiring a non-dispersive soliton solution ($\partial_\xi W_f = \partial_\xi W_s = 0$) gives,
 \eq{\label{e:model_sol}
 [r_2\partial_\eta^2 - \mu ]W_f + W_s W_f^* &= 0, \\
 [ is_1\partial_\eta  + s_2\partial_\eta^2  - 2\mu + \kappa] W_s + \frac{ W_f^2}{2} &= 0.}
 For large phase-mismatch, $\kappa$, neglecting SH dispersion $s_2$, these coupled equations simplify to the nonlinear Sch\"odinger (NLS) equation. With the appropriate combination of FH dispersion and phase-mismatch signs, such that $r_2\kappa < 0$, bright soliton solutions are known to have the form \cite{Buryak2002}: 
 \eq{\label{e:casc_sol}
 W_f(\eta-s_1\xi) &= \pm 2  \mu \sqrt{\alpha} \, \mathrm{sech}(\sqrt{|\mu|}[\eta-s_1\xi]),\\
 W_s(\eta-s_1\xi) &= 2 \mu \, \mathrm{sech}^2(\sqrt{|\mu|} [\eta-s_1\xi]),
 }
 for $r_2=\pm 1$ where $\alpha = [2\mu - \kappa]/\mu \gg 1$. In this limit the FF component is much greater than the SH and the soliton peak power increases with $|\mu|$.

Requiring that any soliton solution must be localised gives the opportunity for further analysis. To meet this requirement both frequency components of the soliton must be exponentially decaying far from the centre. This can be enforced by setting the form of each soliton profile, $W_f = V_f e^{-\theta_f |\eta|}$ and $W_s =V_s e^{-\theta_s |\eta|} $, for $|\eta| \gg 1$. Operating far from the centre of the soliton, $V_{f,s}$ are some small constant amplitudes, therefore Eq. \eqref{e:model_sol} can be linearised. Then requiring that $\theta_f$ and $\theta_s$ have positive real parts ensures exponential decay and yields conditions for localisation of the soliton as,
\begin{align}\label{e:tail_analysis_1}
        4 r_2 \mu &> \nu^2, \\
        \label{e:tail_analysis_2}
    4 s_2 [2\mu - \kappa] &> [\nu - s_1]^2.
\end{align}

While these conditions are required for exponential tails they are not sufficient for soliton existence. Another required condition is the existence of a constant amplitude (CA) solution that acts as the centre of the homoclinic orbit corresponding to the soliton solution. In order to derive a condition for the existence of a CA solution we look at CA solution itself, found by others \cite{Trillo1995} to be
\eq{\label{e:const_amp}
    \Bar{U}_f &= \pm [2\mu \{2\mu - \kappa\}]^{\frac{1}{2}} ,\\
    \Bar{U}_s &= \mu ,
    }
where $U_f = \Bar{U}_f e^{i\mu \xi}$ and $U_s = \Bar{U}_s e^{i[2\mu-\kappa] \xi}$.
As the system (Eq. \eqref{e:model}) is invariant under the transformation
\eq{\label{e:phi_rotation}
(U_f,U_s) \rightarrow (U_f e^{i\phi}, U_s e^{2i\phi}),  \forall \phi
}
 both $U_f$ and $U_s$ can be chosen to be real (at $\xi=0$), and therefore have a fixed phase difference of either $0$ or $\pi$. So the CA solution in Eq. \eqref{e:const_amp} is only valid when it satisfies this fixed phase difference. As $\mu$ is real, $\Bar{U}_s$ will always be real. For $\Bar{U}_f$ to be real however,
\eq{\label{e:CA_requirement}
\mu [2\mu - \kappa]>0
}
is required, giving our condition for the existence of a CA solution.

\begin{figure}[bt!]
    \centering
    \includegraphics[]{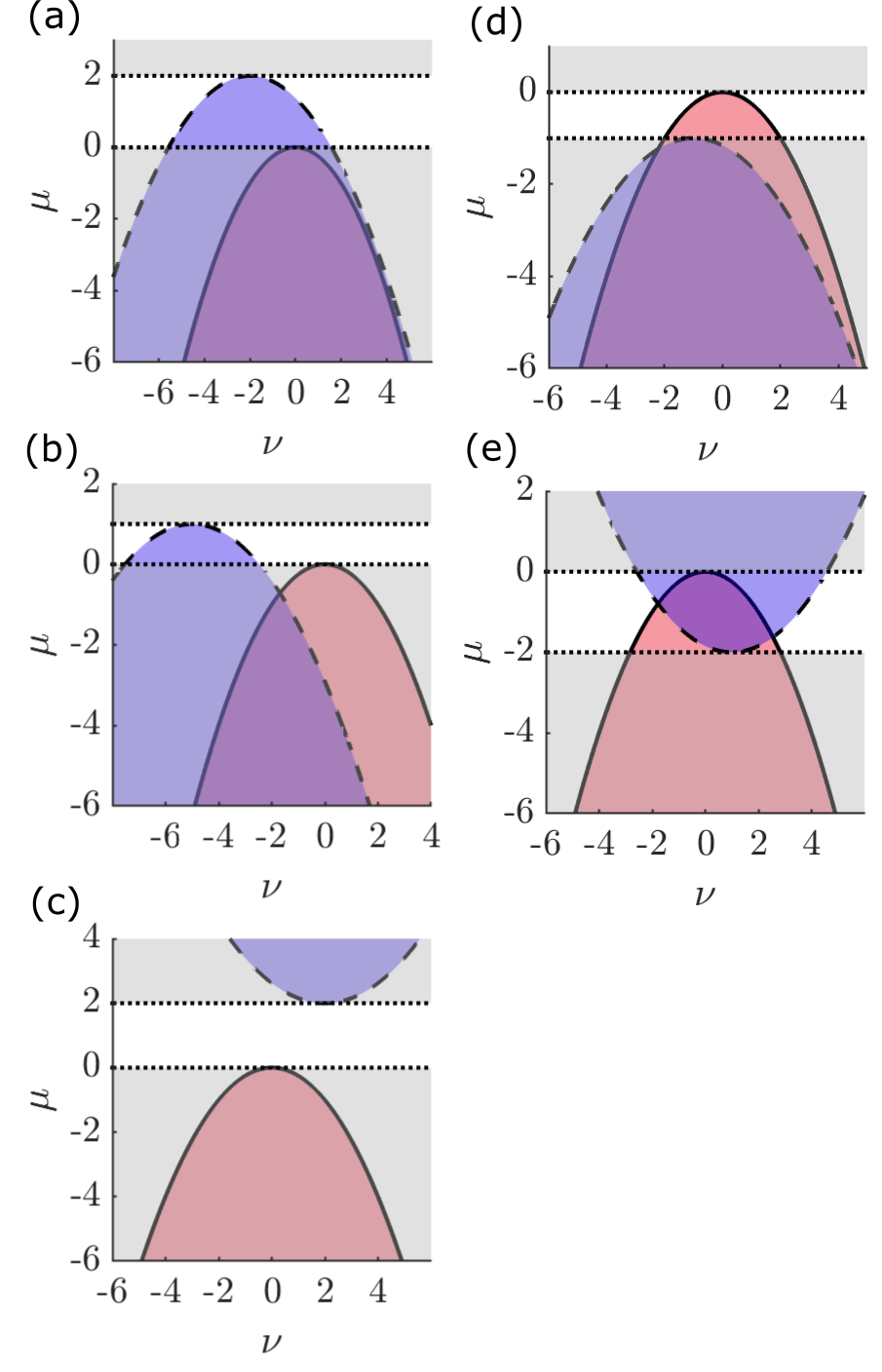}
  	\caption{ Plots of localisation conditions, equations \eqref{e:tail_analysis_1} and \eqref{e:tail_analysis_2}, against soliton parameters $\mu$ and $\nu$. Conditions in equations \eqref{e:tail_analysis_1} and \eqref{e:tail_analysis_2} are marked as shaded regions (in color red and blue respectively) and bounded by solid and dashed lines respectively. Grey shaded regions bounded by dotted lines mark where constant amplitude solutions exist according to the condition in Eq. \eqref{e:CA_requirement}. Parameter $r_2=-1$ throughout and 
  	(a)  $s_2=-0.8$, $\kappa=4$, $s_1=-2$. 
  	(b) $s_2=-0.8$, $\kappa=2$, $s_1=-5$. 
  	(c) $s_2=0.8$, $\kappa=4$, $s_1=2$. 
  	(d) $s_2=-0.8$, $\kappa=-2$, $s_1=-1$. 
  	(e) $s_2=0.8$, $\kappa=-4$, $s_1=1$. }
    \label{f:tail_analysis}
\end{figure}

The conditions for localisation and CA solution existence provide the criteria for localised soliton existence. Graphical representations of these criteria are given in figure \ref{f:tail_analysis} for various waveguide parameters. Each condition on soliton existence is represented by a shaded region in these plots and solitons may exist where all three regions overlap. The five examples given in figure \ref{f:tail_analysis} (a-e) represent distinct regimes of soliton existence. In these examples $r_2=-1$ has been chosen setting normal dispersion in the FF, which is the case for the rest of this work. With normal dispersion in the FH set, solutions in equation \eqref{e:casc_sol} are valid when $\kappa$ is large and positive.

Figure \ref{f:tail_analysis} panels (a - c) show regimes in which the phase-mismatch, $\kappa$, is positive. Conversely panels (d) and (e) show regimes of $\kappa < 0$. In this text these regimes will be referred to by their panel labels (i.e. the regime in figure \ref{f:tail_analysis} panel (a) is regime A).

Regimes A, B and D (shown in figure \ref{f:tail_analysis} (a), (b) and (d) respectively) have areas where all three shaded regions overlap. These regimes are therefore expected to support soliton existence. The difference between these regimes is the range of $|\mu|$ for which solitons are predicted. In regime A solitons are predicted for all negative $\mu$ whereas in regime B solitons are only expected to be possible when $|\mu|$ exceeds a certain threshold value. The key difference between regimes A and B is the shift of the SH localisation condition (Eq. \ref{e:tail_analysis_2}) due to $s_1$. The overlap of FH and SH localisation conditions down to $|\mu| = 0$ is determined by the condition,
\eq{\label{e:no_threshold} \kappa > -s_1^2 / 4s_2 .}
Regime D also exhibits a threshold $|\mu|$ value for soliton existence but in this regime the threshold is set by the CA solution criterion (Eq. \ref{e:CA_requirement}). This threshold is therefore $\mu = \kappa/2$.

It is clear that there are no areas where all three conditions overlap in regime C (shown in figure \ref{f:tail_analysis} (c) ). In this regime the dispersion in the FH and SH have opposite signs. Previous work investigating this regime \cite{Beckwitt2003TemporalHarmonics} has predicted the existence of quasi-solitons where the FF component is localised but the SH has oscillating tails that do not decay far from the centre of the pulse. The existence of these quasi-solitons requires the localisation in the FF and the CA solution to exist but doesn't require localisation in the SH. We therefore expect the existence of these quasi-solitons for all $\mu < 0$ in regime C. It follows that similar quasi-solitons should exist below the threshold in regime B where the condition for SH localisation is not met (but the other two conditions are met).

Finally regime E exists where $\kappa<0 $ and opposite signs of dispersion in the FH and SH. Although clearly similar to regime C, in this regime a threshold for quasi-soliton existence is predicted. Here for negative $\mu$, when $|\mu| < |\kappa/2| $ no solitons or quasi-solitons are expected to exist. However for $|\mu| > |\kappa/2| $ the condition for CA solutions is met but SH localisation is not. Therefore we predicted the existence of quasi-solitons here.

We suggest the term "hard" threshold to describe a threshold set by the CA solution existence criterion \eqref{e:CA_requirement}, as in regimes D and E. Under this threshold no solitons or quasi-solitons are predicted at all. The term "soft" threshold is suggested to describe a threshold set by the SH localisation condition Eq. \eqref{e:tail_analysis_2} as in regime B, where quasi-solitons exist below the threshold.

\subsection{\label{s:disp_waves} Interaction with dispersive waves}

In the previous section we have discussed the existence of solitons and quasi-solitons under the condition of constant GVD (Eq. \eqref{e:dispersion}). Relaxing this condition to allow for the more general case of non-constant GVD, gives the opportunity for further analysis. Consider additional terms such that Eq. \eqref{e:dispersion} becomes,
\eq{
    D_f(\delta) &= - r_2 \delta^2 + C_f(\delta), \\
    D_s(\delta) &= s_1 \delta - s_2 \delta^2 + C_s(\delta) ,
}
where $C_f(\delta)$ and $C_s(\delta)$ are the corrections to the constant GVD of the FF and SH respectively. If these corrections are small the constant GVD soliton solutions should still be approximate solutions. Practically speaking these terms often become relevant close to a zero-GVD point. To analyse this, dispersive waves (DWs) are included into the solution as perturbations of the form,
\eq{\label{e:soliton_pert}
    U_f &= [W_f(\eta) + a_f(\eta, \xi)]e^{i\mu\xi},\\  
    U_s &= [W_s(\eta) + a_s(\eta, \xi)]e^{i(2\mu - \kappa)\xi},
}
where $a_f$ and $a_s$ are small perturbations to the solitons such that $|a_f/W_f| \ll 1 $ and $|a_s/W_s| \ll 1$. Considering these perturbations as a linear combination of, generated (resonant) and existing (pump) DWs, giving
\eq{\label{e:pert}
    a_f &= \psi_f  + p_f e^{i \phi_f(\delta_{f})},\\
    a_s &= \psi_s  + p_s e^{i \phi_s(\delta_{s})} ,
}
where $p_{f,s}$ is the real amplitude of the pump DWs and $\phi_{f}(\delta) = q_{f}(\delta) \xi - \delta_{f} \eta$ and similar for $\phi_s$. $\psi_{f}$ are the generated resonant DWs. In general $a_{f,s}$, $p_{f,s}$ and $\psi_{f,s}$ are all functions of $\eta$ and $\xi$. $\delta_{f,s}$ is the frequency detuning of the pump DWs in the FF and SH respectively and the dispersion in the reference frame of the soliton is
\eq{
    q_f(\delta) &= D_f(\delta) -\mu -\nu\delta, \\
    q_s(\delta) &= D_s(\delta) -2\mu + \kappa -\nu\delta.
}
Substituting equations \eqref{e:soliton_pert} and \eqref{e:pert} into Eq. \eqref{e:model} and taking the Fourier transform gives,
\eq{\label{e:full_FT}
    \left[i\partial_\xi + q_f(\delta)\right] \tilde{\psi}_f &+ [\tilde{W}_s * \tilde{\psi}_f^* + \tilde{W}_f^* * \tilde{\psi}_s] \\&= C_f(\delta) \tilde{W}_f (\delta) \\
    &- \tilde{W_s}(\delta) * \tilde{p}_f (\delta) e^{-i q_f(\delta_f)\xi }  \\ 
    &- \tilde{W}_f^*(-\delta) *  \tilde{p}_s (\delta)  e^{i q_s(\delta_s)\xi }, \\
    \left[i\partial_\xi + q_s(\delta)\right] \tilde{\psi}_s  &+ [\tilde{W}_f * \tilde{\psi}_f] \\&=  C_s(\delta) \tilde{W}_s \\
    &- \tilde{W}_f  * \tilde{p}_f(\delta) e^{i q_f(\delta_f)\xi } 
}
where $\tilde{f}(\delta)$ denotes the Fourier transform of any function $f(t)$ and $f(t)*g(t)$ denotes a convolution of functions $f(t)$ and g(t). The left-hand side (LHS) acts as an oscillator driven by the terms on the right-hand side (RHS). As $\tilde{\psi}_{f,s}$  and $\tilde{W}_{s,f}$ are all localised functions their convolution (second terms on LHS of equations \eqref{e:full_FT}) will only make small contributions to the resonant frequencies of the system. For this reason these terms are neglected in the following analysis.

\subsubsection{\label{s:dw_cherenkov} Cherenkov radiation}

In the case of a system initially free from dispersive waves ($p_f=p_s=0$), Eq. \eqref{e:full_FT} simplifies to the driven oscillator equations,
\eq{
    \left[i\partial_\xi + q_f(\delta)\right]\tilde{\psi}_f &=  C_f(\delta) \tilde{W}_f, \\
    \left[i\partial_\xi + q_s(\delta)\right]\tilde{\psi}_s &= C_s(\delta) \tilde{W}_s.
}

Taking $\psi_{f}$ of the form $\psi_{f} \propto e^{iq_{f}\xi - i \delta \tau}$ (and similar for $\psi_s$) and matching wavenumbers between the oscillating and driving terms gives the resonant conditions
\eq{\label{e:cherenkov_conditions}
    q_f(\delta) = 0 , \\
    q_s(\delta) = 0 .
}

Where these conditions hold true the soliton is resonant with dispersive waves in the system. Radiation with the wave vector, $q_{f,s}$ (and corresponding frequency detuning, $\delta$) satisfying these conditions will be emitted from the soliton. This process is known as resonant or Cherenkov radiation \cite{Skryabin2005TheoryFibers}. In the previous section solitons were found to be possible in regimes A, B and D. With corrections to the constant GVD ($C_f(\delta)$ and $C_s(\delta)$), these solitons become resonant with dispersive waves in the system and emit Cherenkov radiation. As they are no longer localised the solitons become quasi-solitons.

Similar analysis can be applied to the quasi-solitons in regimes B, C and E. We find that SH dispersion ($s_1$ and $s_2$) acts as the perturbation in this case. With these terms set to zero a soliton solution exists. Reinstating these terms produces resonances between the soliton and dispersive waves in the system and quasi-solitons are predicted. At these resonances Cherenkov radiation is expected, this can also be interpreted as the oscillating tails of the soliton. The addition of further dispersive terms $C_f(\delta)$ and $C_s(\delta)$ may shift the frequency of this Cherenkov radiation.

\subsubsection{\label{s:dw_pump} Pumped radiation}

Any waves in the system that are not part of the soliton may perturb the soliton leading to emission of DWs at new frequencies. Here they are referred to as pump DWs, $p_f$ and $p_s$, in the FF and SH respectively. This includes any deliberately introduced pump into the system or radiation previously emitted by the soliton in question or other solitons in the system. In the following analysis pump DWs are assumed to be continuous wave and are therefore delta functions in frequency. This allows their convolutions to be evaluated simply giving,
\eq{
    \left[ i\partial_\xi + q_f(\delta) \right] \tilde{\psi}_f = 
    &- p_f \tilde{W}_s(-\delta_f)  e^{-i q_f(\delta_f)\xi }\\
    &- p_s [\tilde{W}_f(-\delta_s)]^* e^{i q_s(\delta_s)\xi }, \\
    \left[i\partial_\xi + q_s(\delta) \right] \tilde{\psi}_s = 
    &-  p_f \tilde{W}_f(\delta_f)  e^{i q_f(\delta_f)\xi },
} 

with the resonant conditions,

\eq{\label{e:SP_conditions}
    q_f(\delta) &= -q_f(\delta_f) , \\
    q_f(\delta) &= q_s(\delta_s) , \\
    q_s(\delta) &= q_f(\delta_f) .
}

This radiation is produced by the interaction of the soliton and the pump DWs, and will therefore be referred to as pumped radiation. For a pump DW with a small detuning from the FF central frequency two resonant conditions exist, one in the FF and one in the SH. A pump DW with a small detuning from the SH central frequency, has one possible resonance in the FF.

Considering the specific case of previously emitted Cherenkov radiation, where $q_f(\delta_f) = 0$ or $q_s(\delta_s) = 0$. Substitution into Eq. \eqref{e:SP_conditions} shows that, the Cherenkov conditions from Eq. \eqref{e:cherenkov_conditions} are reproduced. This shows that previously emitted Cherenkov radiation can not produce new frequencies when interacting with the soliton it was emitted from. In general, Cherenkov radiation emitted by one soliton interacting with a different soliton (with different central frequency, $\mu$ or $\nu$) would produce new frequencies.

\section{ Solitons in nano-waveguides }
\subsection{ Waveguide Simulation }
\label{s:wg_simulation}

In the previous section we have seen that existence of soliton and quasi-soliton solutions depends strongly on the waveguide parameters. In this section we present simulated data for two waveguide structures. These data are analysed from the point of view of soliton and quasi-soliton existence as discussed in the previous section. This section is intended to clearly show how the generic theory presented so far maps onto specific waveguide geometries that can be experimentally realised.

\begin{figure*}[bth!]
    \centering    
    \includegraphics[]{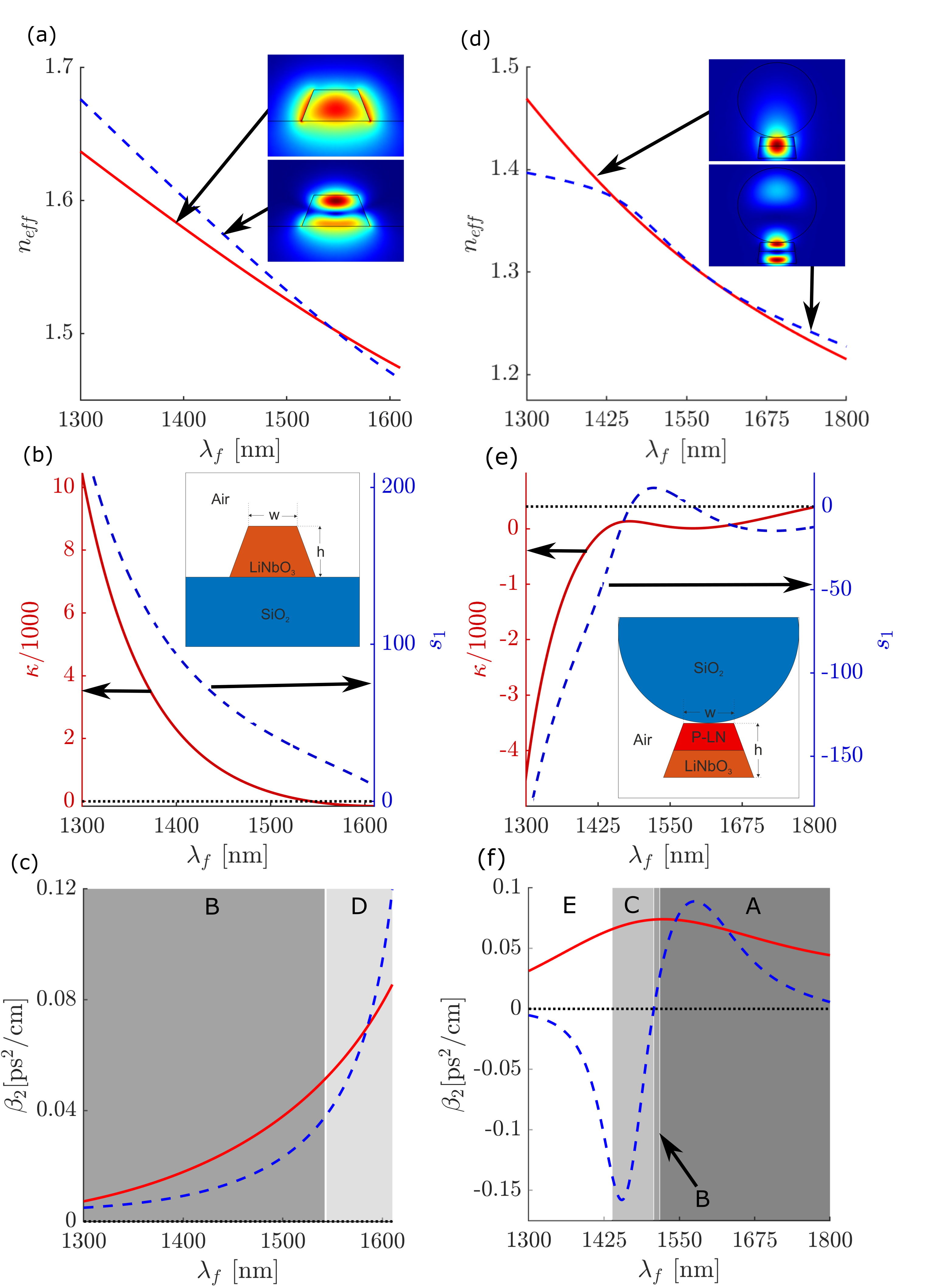}
    
    \caption{ For Lithium Niobate on insulator (LNOI), $n_{e\!f\!f}$ (a), $\kappa$ and $s_1$ (b) and $\beta_2$ (c) plotted against the FF wavelength $\lambda_f$. Insets in (a) show the FF (upper) and SH (lower) mode profiles. Inset in (b) shows the LNOI waveguide cross-section, $h$ and $w$ label the height and width of the waveguide respectively. Data shown is for LNOI structure with height of $350$nm and width of $500$nm. 
    For free standing Lithium Niobate and microfibre (hybrid) structure, $n_{e\!f\!f}$ (d), $\kappa$ and $s_1$ (e) and $\beta_2$ (f). Insets in (d) show the FF (upper) and SH (lower) mode profiles. Inset in (e) shows the cross-section of the hybrid waveguide, $h$ and $w$ label the height and width of the waveguide respectively. Data shown is for a waveguide height of $300$nm, width of $470$nm and a microfibre diameter of $1100$nm. Solid and dashed curves (red and blue in colour) are for FF and SH modes respectively. Dotted black lines mark $\kappa = 0$, $s_1=0$ and $\beta_2 = 0$. Shaded regions mark the different soliton supporting regimes labeled (A-E). }
    \label{f:neff_dispersion}
\end{figure*}

Figures \ref{f:neff_dispersion} (a-c) present simulated data for a Lithium Niobate on Insulator (LNOI) structure \cite{Poberaj2012LithiumDevices, Yu2019CoherentWaveguides, Luo2018HighlyWaveguide}. The structure of this waveguide is shown in the inset of figure \ref{f:neff_dispersion} (b) and consists of a ridge of Lithium Niobate on a Silica substrate. Figure \ref{f:neff_dispersion} (a) shows the $n_{e\!f\!f}$ data for simulated FF and SH modes in the structure. The insets show the transverse mode profiles of the chosen modes. A fundamental mode was chosen for the FF where a higher order mode was selected for the SH, this allows for phase-matching between the modes. Phase-matching occurs when both modes have the same propagation constant $\beta(\omega)$ and from Eq. \eqref{e:beta_neff}, the same effective index, $n_{e\!f\!f}$. The phase-mismatch parameter, $\kappa$, is shown in figure \ref{f:neff_dispersion} (b) and is zero at phase-matching. It is clear that in this waveguide phase-matching occurs at around 1550nm.

Another important waveguide parameter to consider is the group velocity mismatch parameter, $s_1$. This is plotted in figure \ref{f:neff_dispersion} (b). In this waveguide it is clear that $s_1$ rapidly decreases as wavelength increases but does not reach zero for the wavelengths shown. Figure \ref{f:neff_dispersion} (c) shows the GVD parameters $\beta_2$ for both FF and SH modes. Both modes show slowly varying GVD and with no zero-GVD points, the GVD of both modes remains normal for the wavelength range shown. The regimes of soliton existence outlined in the previous chapter are indicated in figure \ref{f:neff_dispersion} (c) by shaded areas and labeled with their corresponding letters. This waveguide supports regimes B and D for a broad range of wavelengths. From this we would expect solitons to be possible in this waveguide structure but with "soft" or "hard" thresholds depending on the regime (determined by wavelength).

Simulated data for the hybrid waveguide structure is shown in figures \ref{f:neff_dispersion} (d-f). This hybrid structure consists of a suspended Lithium Niobate core and a silica microfibre, the cross-section is given as an inset in figure \ref{f:neff_dispersion} (e). More details about this structure can be found in previous work  \cite{Gorbach2015Microfiber-LithiumChip,Main2016HybridSources,Cai2018HighlyNano-waveguides}. Figure \ref{f:neff_dispersion} (d) shows simulated $n_{e\!f\!f}$ data for FF and SH modes in this waveguide. Insets are included showing the transverse mode profiles. The phase-mismatch and group velocity mismatch parameters are plotted in figure \ref{f:neff_dispersion} (e). The GVD parameters for both modes are plotted in figure \ref{f:neff_dispersion} (f).

Comparison with the data from the LNOI structure shows clear differences. The hybrid structure shows a broad wavelength range where both $\kappa$ and $s_1$ are low. The GVD parameter for the SH also changed much more rapidly in the hybrid structure, exhibiting a zero-GVD point near the centre of the wavelength range shown here. This zero-GVD point results in a change of sign of the SH dispersion making regimes in which FF and SH have opposite signs of dispersion available. Again the soliton regimes in this structure are plotted as shaded areas in figure \ref{f:neff_dispersion} (f). This structure provides 3 broad wavelength ranges where regimes A, C and E exist. Therefore we would expect to find solitons  and quasi-solitons without any threshold in regimes A and C respectively. Quasi-solitons with a "hard" threshold are expected in regime E.
With these two waveguide structures are therefore expected to exhibit all five of the predicted soliton regimes.

\subsection{Pulse propagation}
\label{pulse_propgation}

Using the data for the example nano-waveguides we can simulate the propagation of pulses in these structures using the Split-Step Fourier method. In these simulations dispersion has been taken as a Taylor expansion truncated to third-order such that the corrections to constant GVD are $C_f(\delta) = -r_3 \delta^3$ and $C_s = -s_3 \delta^3$. This approximation of dispersion as truncated Taylor expansion is accurate close to the pulse central frequencies and allows the effects of correction terms $C_f$ and $C_s$ to be demonstrated. 

In this section we present XFROG (cross-correlation frequency-resolved optical gating) spectrograms of pulses after propagation. An XFROG spectrogram is a well-known method for resolving both temporal and spectral features of a pulse \cite{Skryabin2005TheoryFibers, Skryabin2010ColloquiumSupercontinuum}. Here the XFROGs produced with the numerical integration of
\eq{
I(t, \omega) = ln \left| \int^{+\infty}_{-\infty} dt' A_{ref}(t' - t) U(t') e^{-i \omega t'}\right|,
}
where $A_{ref}$ is the envelope of a Gaussian reference pulse, and $U$ is either the FH or SH field envelope. The plots for the analytic predictions of Cherenkov radiation are included for comparison using $\mu$ values estimated from the simulations. In all the simulations present here the emitted Cherenkov radiation matches closely with the resonance predictions.

\begin{figure*}[!bth]
    \includegraphics[]{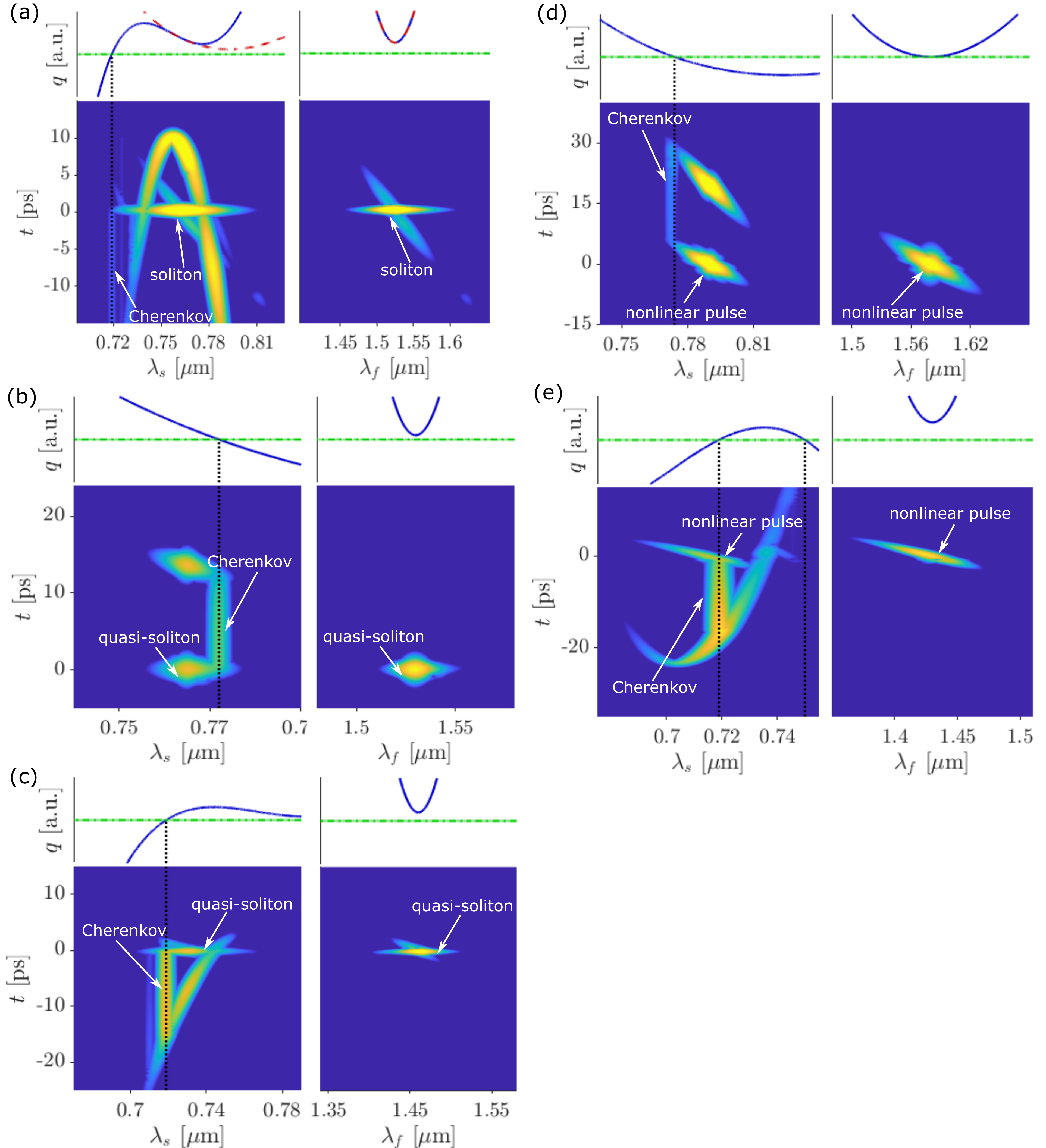}
    \caption{ XFROG plots of pulses after simulated propagation in nano-waveguide structures. Each panel consists of FH (right) and SH (left). Resonance predictions are included above XFROG plots for comparison. Wavevectors of dispersive waves are shown as solid lines (blue and red in colour, truncated to third and second order dispersion respectively). Soliton wavevectors are shown as dot-dashed line (green in colour). Resonances occur where these lines intersect, and are marked by vertical dotted lines. Panels (a), (c) and (e) are in the hybrid structure, panels (b) and (d) are in the LNOI structure. 
    Input pulse parameters: 
    (a) 1.1kW peak power, 140fs duration, central wavelength 1525nm, propagation distance 20mm. 
    (b) 1.9kW peak power, 560fs duration, central wavelength 1530nm, propagation distance 10mm. 
    (c) 760W peak power, 180fs duration, central wavelength 1460nm, propagation distance 10mm. 
    (d) 14kW peak power, 560fs duration, central wavelength 1580nm, propagation distance 30mm. 
    (e) 1.2kW peak power, 120fs duration, central wavelength 1430nm, propagation distance 10mm.
    }
    \label{f:regimes_XFROG}
\end{figure*}

In figure \ref{f:regimes_XFROG} the waveguides and wavelengths have been chosen such that each panel (a-e) represents each regime (A-E). The input pulse for each simulation was set as in Eq. \eqref{e:casc_sol} for the FF field and zero in the SH field.

Figure \ref{f:regimes_XFROG} (a) shows a soliton formed after propagation of a FF pulse in regime A in the hybrid structure. Although the input pulse used to generate the soliton here had a peak power of 1.1kW we found soliton generation was possible for all peak powers attempted (lowest attempted was 15W peak power). The resonance predictions in this panel shows both the dispersion truncated to second order and third order. As expected for a localised soliton there are no resonances present when dispersion is truncated to second order. With third order dispersion added there is a single resonance predicted far from the centre of the soliton where low intensity Cherenkov radiation is observed.

Figures \ref{f:regimes_XFROG} (b) and (c) show quasi-solitons formed in the LNOI and hybrid structures respectively. In both cases soliton component in the FF is localised but the SH is strongly emitting Cherenkov radiation. The wavelength of this radiation matches closely with that expected from the resonance predictions. This type of quasi-soliton is predicted below the threshold in regime B and for all powers in regime C. At high powers solitons localised in both FF and SH are expected in regime B but due to the large walk-off ($s_1$) in the LNOI structure this threshold is estimated to be 500MW peak power and therefore experimentally unattainable. Both these quasi-solitons are expected to emit Cherenkov radiation without the inclusion of third order dispersion which was verified. With the addition of third-order dispersion the wavelength of the resonances were shifted and in the case of regime C one resonance was removed entirely leaving the quasi-soliton shown in figure \ref{f:regimes_XFROG} (c).

Figures \ref{f:regimes_XFROG} (d) and (e) show pulses after propagation in regimes D and E respectively. Both of these regimes exhibit a "hard" threshold under which no solitons or quasi-solitons are expected. In simulations in both of these regimes we were unable to find either solitons or quasi-solitons at low powers. Typically we found that a nonlinear pulse would form having components in the FF and SH. This pulse would exhibit some aspects of a soliton and quasi-soliton such as Cherenkov radiation and the SH component staying locked to the FF component. These nonlinear pulses are however not solitons as they disperse as they propagate. Non-solitonic pulses emitting resonant radiation has been predicted before in previous work \cite{Webb2013GeneralizedOptics, Roger2013High-energyRegime}. In these regimes we did not observe formation of solitons at low powers even with the addition of SH pulses in the initial conditions.

\begin{figure*}[!bth]
    \centering
    \includegraphics[]{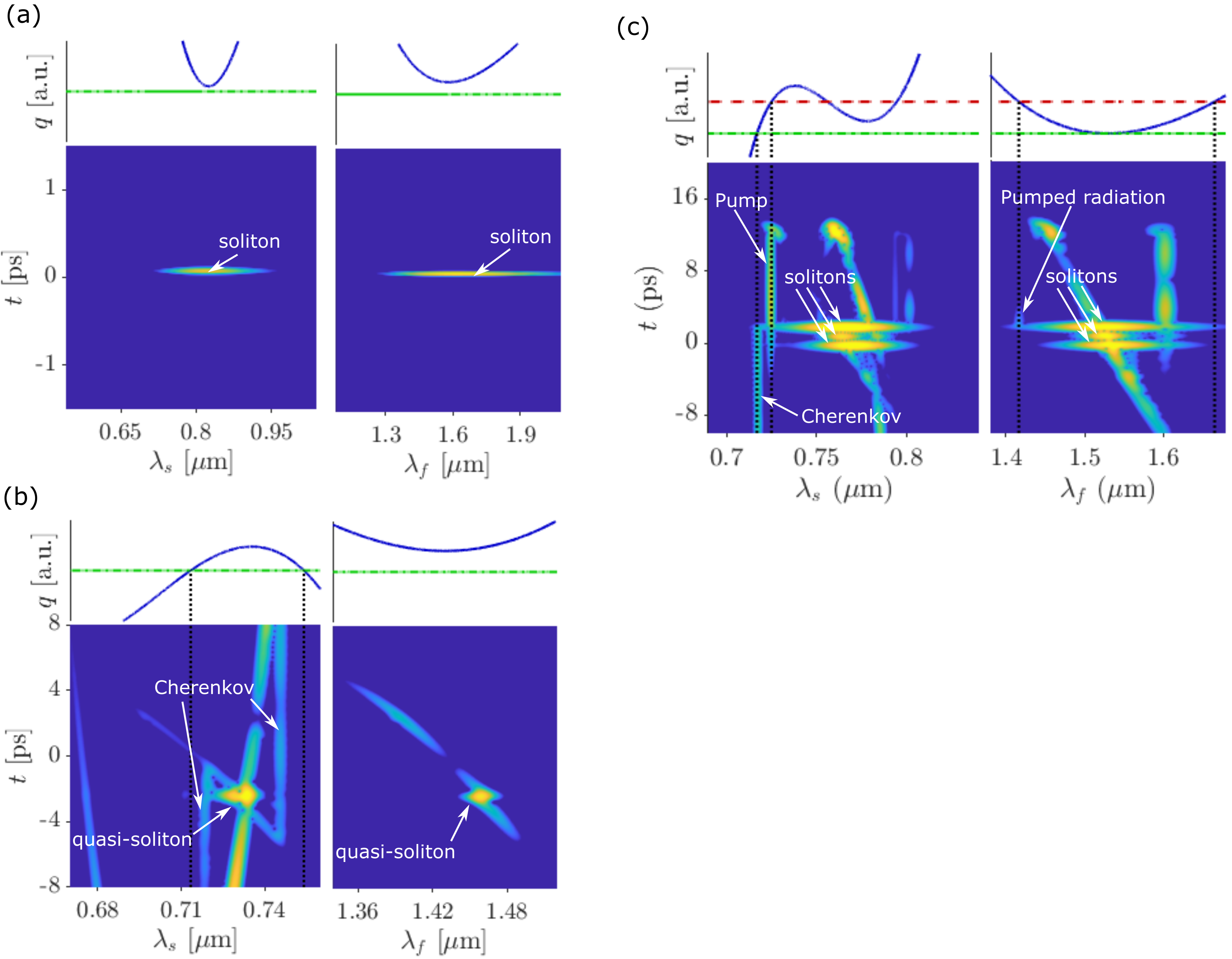}
    \caption{ \label{f:extra_XFROG} XFROG plots of pulses after simulated propagation in nano-waveguide structures. Each panel consists of FH (right) and SH (left). Resonance predictions are included above XFROG plots for comparison. Wavevectors of dispersive waves are shown as solid lines (blue in colour, truncated to third order dispersion). Soliton wavevectors are shown as dot-dashed line (green in colour). Dashed lines (red in color) show the wavevectors of the soliton interactive with pump frequencies. Resonances occur where these lines intersect. Vertical dotted lines mark the resonance wavelengths. Panels (b) and (c) are in the hybrid structure, panel (a) is in the LNOI structure.
    Initial pulse parameters: 
    (a) 6.3MW and 19MW peak power in the FH and SH respectively, 16fs duration, central wavelength 1580nm, propagation distance 10mm. 
    (b) 17kW and 21kW peak power in the FH and SH respectively, 25fs duration, central wavelength 1430nm, propagation distance 15mm. 
    (c) 2.7kW peak power, 390fs duration, central wavelength 1530nm, propagation distance 26mm.
    }
\end{figure*}

Figure \ref{f:extra_XFROG} (a) shows a soliton after propagation in regime D. The initial conditions for this simulation was a numerically calculated soliton solution. This was necessary as the analytical solution (Eq. \eqref{e:casc_sol}) is no longer a good approximate solution in this regime. This result shows that for extremely high powers solitons can exist in regime D. This high power is set by the hard threshold in this region.

It has not been possible to generate a quasi-soliton in regime E. Attempts to excite quasi-solitons in this regime has resulted in quasi-solitons with a different central frequency, in a range that isn't in regime E. An example of this type of soliton is shown in figure \ref{f:extra_XFROG} (b). We expect that exciting solitons in this regime is particularly difficult due to strong Cherenkov radiation expected from quasi-solitons in this regime destabilising any potential solitons. The near by existence of different regimes where soliton existence was more favourable is also expected to make soliton generation here more challenging.

Figure \ref{f:extra_XFROG} (c) shows the result after propagation of a broad higher power pulse in regime A. The input field in the FF was of the form of the analytic soliton solutions (Eq. \eqref{e:casc_sol}) increased by a scale factor and zero in the SH. Multiple solitons are visible near $t=0$ with one of them emitting Cherenkov radiation. A soliton/pulse at around $t=12$ps can be seen emitting its own Cherenkov radiation which is labeled as 'pump'. This radiation interacts with one of the solitons near $t=0$, perturbing it and acting as a pump for further radiation. This further radiation can be seen labeled with 'pumped radiation', and coincides with the resonance predictions. We can be sure that this pumped radiation is distinct from Cherenkov radiation as it begins to be formed as the soliton begins to interact with the pump waves.

 Previous literature in this area suggests peak input powers of a few kW \cite{Yu2019CoherentWaveguides} are experimentally achievable for similar waveguides and wavelengths. This suggests that it should be experimentally possible to generate many of the solitons and quasi-solitons observed here. It is important to highlight that the solitons predicted and simulated in figure \ref{f:regimes_XFROG} (a) experience normal GVD ($\beta_2 > 0$) in both the FH and SH. As Kerr solitons cannot exist for normal GVD \cite{Buryak2002} any experimentally observed solitons in this frequency range could only be due to $\chi^{(2)}$ nonlinearity.

\section{Conclusion}

The existence of temporal solitons in generic $\chi^{(2)}$ waveguides has been investigated. Conditions for the existence of localised solitons have been analysed and five distinct regimes of soliton and quasi-soliton existence have been identified. Two nano-waveguide structures were simulated and all five regimes were found to exist for different wavelengths in these two waveguides. Predictions for soliton and quasi-soliton existence have been confirmed by pulse propagation simulation in the proposed nano-waveguide structures. 

The interaction of solitons with higher order dispersion terms has been predicted to produce Cherenkov radiation from both FH and SH soliton components. Soliton propagation simulations have confirmed these predictions.
Radiation due to dispersive wave pumps has also been predicted and simulated. The wavelengths of this radiation in simulation coincides closely with those expected from analytic predictions. The generation of multiple solitons and quasi-solitons under opposite signs of dispersion has also been simulated for experimentally attainable peak powers.

This model is intended to provide a useful theoretical basis for low-power soliton generation in $\chi^{(2)}$ waveguides. We hope that our analysis provides the possibility of optimising soliton-assisted frequency conversion in Lithium Niobate nano-waveguides.

\section{Acknowledgements}
WR acknowledges funding and support from the Engineering and Physical Sciences Research Council (EPSRC) Centre for Doctoral Training in Condensed Matter Physics (CDTCMP), Grant No. EP/L015544/1.

\bibliography{mendeley.bib}

\end{document}